**Recent Progress in Pencil Beam Scanning FLASH Proton Therapy: A Narrative Review**

Shouyi Wei[1], Chengyu Shi[1], Chin-Cheng Chen[1], Sheng Huang[1], Robert H. Press[1], J. Isabelle Choi[1,2], Charles B. Simone, II[1,2], Haibo Lin[1,2], Minglei Kang[1*]

[1]New York Proton Center, New York, NY 10035

[2]Department of Radiation Oncology, Memorial Sloan Kettering Cancer Center, New York, New York

* Author to whom any correspondence should be addressed

Email: mkang@nyproton.com

## Abstract

**Background and Objective**: Recent experimental studies using ultra-high dose rate radiation therapy (FLASH-RT) have shown improved normal tissue sparing and comparable tumor control compared to conventional dose rate RT. Pencil beam scanning (PBS) proton therapy with superior dosimetry characteristics has begun to draw attention to the delivery of conformal FLASH-RT for preclinical studies. This review aims to provide recent updates on the development of PBS FLASH-RT.

**Method**s: The information summarized in this review article is based on search results in databases such as PubMed and search engines like Google Scholar, with keywords including pencil beam scanning, proton therapy, proton FLASH, Bragg peak FLASH, etc., with English articles from the year of 2014-2022.

**Content and** Findings: This review summarizes of recent developments in PBS FLASH proton therapy (FLASH-PT), including PBS dose rate characterization, current delivery limitations, treatment planning, and biological investigations.

**Conclusions:** As PBS FLASH delivery has enabled successful biological studies using transmission beams, the further improvement in PBS Bragg peak FLASH technologies will result in more advanced treatment plans associated with potentially improved outcomes.

## Introduction

FLASH radiotherapy (RT) is a promising radiation therapy modality that has shown reduced normal tissue toxicity and isoeffective tumor control in many in-vivo studies (1-11), compared to conventional dose rate RT. These studies have shown preserved functionality in various sites, including the lung (1), skin (3), brain (2, 7), and abdomen (5,6,8,9,10). The first human study was reported on a CD30+ T-cell cutaneous lymphoma patient treated with FLASH electron beams, and the treatment achieved both promising normal skin protection and tumor response (11).

The majority of FLASH studies were based on electron beams from modified linear accelerators. Electron beam therapy has limitations in treating deep-seated targets and achieving sufficient field dose conformity. Such challenges can be conveniently addressed by proton therapy that offers variable therapeutic depths and rapid dose fall-offs beyond Bragg peaks, leading to superior dose conformity (12). For current proton PBS planning and treatment, multiple energy layers are used to generate spread-out Bragg peaks (SOBP) to cover the target volume. However, using the conventional SOBP becomes difficult for FLASH-RT to deliver ultra-high dose rate spots across an entire target volume due to the beam transmission efficiency, i.e., the beam current dropped dramatically for lower energies beams in the case of energy degradation-based cyclotron systems (13-14). Additionally, the typical energy/layer switch time is ~200 ms for energy degradation-based cyclotron systems (15) and on a scale of >1000 ms for synchrotron systems (16) with increased delivery time. Therefore, the current PBS planning strategies using multiple energies hardly reach the FLASH dose rate threshold in OARs (17). To use SOBPs for conformal FLASH application, researchers used rotation modulators or ridge filters to create SOBPs based on scattering proton systems. Kourkafas et al. used a single scattering system with a rotating modulator wheel to create a proton SOBP at FLASH dose rates of approximately 75 Gy/s based on a high-intensity 68 MeV cyclotron (18). A collimating aperture shapes the radiation field at the isocenter to enable the treatment for an extremely small field, i.e., ocular irradiation (18). Ridge filters are proton beam-forming devices that provide a simultaneous option for multiple energy extractions (19-20). They are stationary devices placed along the beamline; thus, they require no added modulation other than the rotating modulator (19). The scattering proton systems combined with range modulators or ridge filters provide a viable solution to the proton FLASH. Kim et al. (19) and Evans et al. (21) conducted small animal studies using ridge filters to create SOBPs based on clinical cyclotron and synchrocyclotron systems.

Compared to the scattering-based techniques, pencil beam scanning (PBS) is the state-of-the-art proton delivery technique and is the dominant treatment modality in proton therapy. It uses pencil beam proton spots to deliver the field dose instead of using a uniform scattered proton beam; therefore, it can achieve superior dose conformity to the scattering technique for clinical targets of variable sizes. The transmission method using PBS beam shoot-through from different angles with a single high-energy is more practical to reach the FLASH dose rate and also minimizes range uncertainties in heterogeneous tissues (22). Several research groups studied the dosimetry and dose rate performance to the head and neck (H&N) (17,23-24) and lung cancers (22, 25-26) using transmission proton FLASH beams. The first human clinical trial was initialized to study the safety and feasibility of using transmission FLASH beams to treat bone metastases (27).

While transmission proton beams still deposit significant doses to OARs or normal tissues beyond the targets, a novel Bragg peak FLASH method was developed to deliver conformal doses using single-energy PBS beams to take advantage of proton beams. By pulling back the ranges of the highest energy proton beams using beam-specific range pullbacks that consist of a target-specific range compensator and a universal range shifter to adapt to the target distally, the exit dose of proton beams can be eliminated to better protect OARs, while still preserving FLASH dose rate delivery by meeting the minimum MU constraints with a spot map optimization and inverse treatment planning capability offered by an in-house TPS (28). Thus, the review will focus on the recent progress of proton PBS FLASH-RT. We present the following article in accordance with the narrative review reporting checklist.

We present the following article in accordance with the Narrative Review reporting checklist (available at https://tro.amegroups.com/article/view/10.21037/tro-22-1/rc).

**Methods**

All the information gathered in this review article is based on search results in databases such as PubMed and/or search engines like Google Scholar, with keywords including proton therapy, FLASH, proton FLASH, FLASH-RT, pencil beam scanning, etc. We only include articles published in English from the year 2014 to the year 2022.

**Table 1. The search strategy summary**

| Items | Specification |
| --- | --- |
| Date of Search | 2022/1/31 |
| Databases and other sources searched | PubMed, google scholar |
| Search terms used | Pencil beam scanning, Proton therapy, Proton FLASH, Bragg peak FLASH, |
| Timeframe | 2014-2022 |
| Inclusion and exclusion criteria | Inclusion: research article, English |
| Exclusion: none | |
| Selection process | S. Wei and M. Kang conducted the selection, and it was conducted independently with consensus obtained throughout the authors in the author list. |

**Current machine capacity in delivery FLASH-PT**

Challenge in PBS FLASH-PT delivery mainly comes from practical machine-related factors, including the maximum available beam currents in the treatment room, the spot delivery time, and scanning speed. Besides, multiple energy layers employed in the conventional dose rate PBS PT cannot be readily used currently in PBS FLASH-PT due to the inefficient energy switching in existing proton treatment systems (29). The single-energy transmission beams have been currently employed for PBS FLASH-PT, which is able to provide sufficient beam current to reach ultrahigh FLASH dose rate, which is estimated to be over 40 Gy/s without any significant hardware upgrade based on the current existing clinical systems (22). Diffenderfer et al. (5) reported a maximum 350 nA beam current for FLASH based on an IBA cyclotron system. Nesteruk et al. (30) reported their experience with the FLASH commissioning based on a cyclotron that achieved > 680 nA beam current at the isocenter. As the MU definition varies amongst different vendors, the beam current at the treatment room is a universal quantity to compare the dose rate between different machines. In order to describe the dose rate precisely, the correlation between beam current and the number of protons per MU needs to be well established by Monte Carlo simulation or experimental methods (31). Different types of machines will have different spot delivery mechanisms, and awareness of the delivery differences between different machines will be important to model the dose rate correctly. For example, the Varian ProBeam system works under a layer-wise delivery manner, meaning that the spot dose rate of each layer is determined by the minimum MU/spot. A spot peak dose rate (SPDR) was defined as the maximum dose rate at the central axis in the water phantom (25) to compare different dose rates among different machines. Figure 1(a) shows the dose rate distribution for a single spot with 100 MU in a water phantom. Figure 1(b) displays the correlations of SPDR, nozzle beam current, and MU/s for the FLASH mode. Given a beam current of 215 nA, this will enable the 500 MU/spot to be delivered within 2 ms (22).

Another major challenge with the delivery of PBS FLASH proton beams comes from the accurate monitoring of the absolute dose delivered, as, at high beam current, the recombination effect increases dramatically, which approaches up to a 20% efficiency drop at the highest cyclotron current of ~ 800 nA (32). To address this challenge, either reliable correction methods or dose-rate independent detectors will be necessitated to reliably control and monitor the dose delivered for PBS FLASH PT. Some detectors (33, 34, 35) indicating low dose rate dependence may also be affected by the temporal drops, noise in the signal, and long integration time, which also poses a challenge in FLASH delivery.

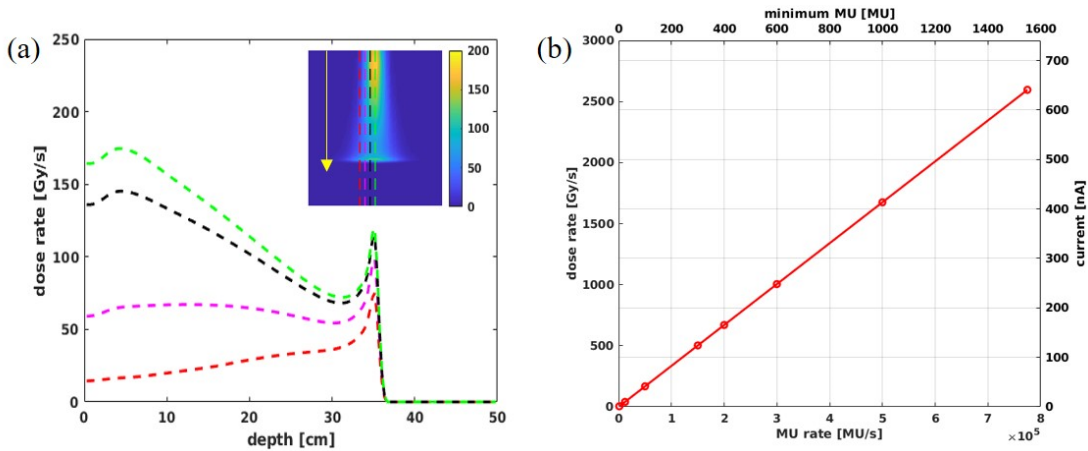

Figure 1 (a) The dose rate profile for a 240 MeV single spot with 100 MU at the central plane along depth direction (indicated by the arrow), (b) the theoretical calculation for nozzle current, minimum MU/spot, SPDR, and MU for proton beam under FLASH mode.

**Characterization of the PBS Dose rate for FLASH-PT**

Currently, there is no consensus on the definition of the PBS proton therapy dose rate calculation for normal tissues and targets (22). With the dose rate as a deciding factor, it is natural to envision the FLASH treatment planning should encompass the dose rate constraints for organ-at-risks (OARs), similar to dose constraints in conventional treatment planning. A PBS field uses scanning pencil beam spots to deliver doses, for a single proton spot, dose rate distribution will follow the same spatial dose distribution but scaled by a time denominator. A PBS field needs hundreds of pencil beam spots to be delivered sequentially, and thus the dose rate for a region of interest (ROI) or voxel must be calculated from the neighboring scanning spots. Figure 2 shows different dose contributions to a voxel in the field of 5 x 5 $cm^2$ from different adjacent spots simulated with a spot sigma of 3.5 mm. The star represents an arbitrary ROI, and the number shows the dose weighting to the ROI from the neighboring spots. Therefore, the field dose rate quantification greatly differs from the electron and proton scattering techniques that simultaneously deliver uniform fluence to the entire field. In addition, the switch time between the spots may further complicate the dose rate quantification. Several research groups have proposed different ways of quantifying dose rate in PBS plans, emphasizing a subset of dose delivery features based upon different assumptions.

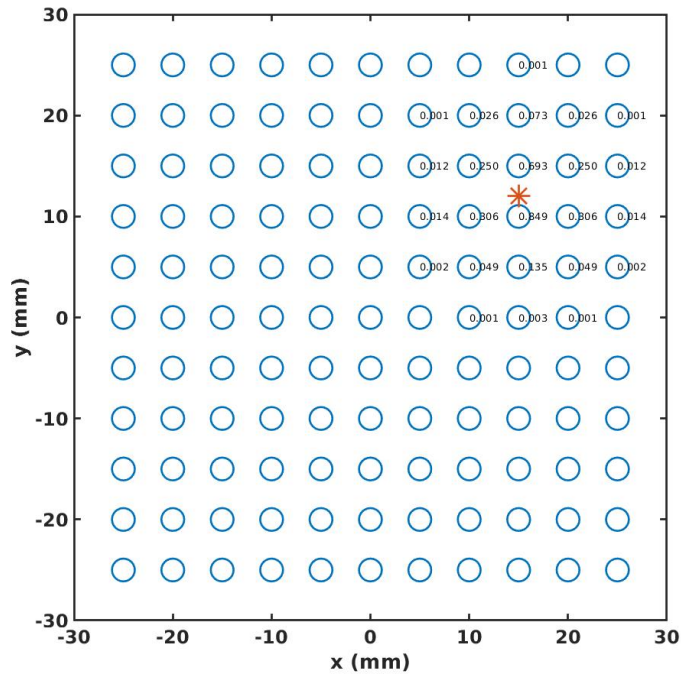

Figure 2 Dose or dose rate contribution from different spots to a voxel in a 5x5 cm$^2$ spot map with spot sigma of 3.5 mm. The circle represents the spots, the star represents an arbitrary voxel or region of interest (ROI), and the number shows the dose weighting to the ROI from the neighboring spots.

van de Water et al. (17) defined the dose-averaged dose rate (DADR) as the summed spot dose rate normalized by spot dose weighting factors to a particular voxel in the irradiated volume. While this metric directly quantifies the volume dose rate distribution, it essentially ignores the spot and spot scanning time in PBS. Gao et al. (26) developed a treatment planning framework that optimizes the dose rate coverage using DADR and plan quality by adding the minimum MU constraint in the FLASH treatment planning. Folkerts et al. (36) proposed the average dose rate (ADR)-the accumulated dose ($D_j$-2d*) divided by dose accumulation time $T_j$. For a particular voxel j, ($D_j$-2d*) is the total dose deposited in voxel j during the irradiation $T_j$, d* is a preset dose threshold that determines the irradiation start time $t_0$ and the end time $t_1$. By applying the dose threshold, d*, the nonsignificant dose accumulation to voxel j from all the scanning spots is excluded from the dose rate calculation. ADR typically yields smaller values than DADR, which is also sensitive to the dose threshold chosen in the method (22). As there is more evidence suggesting there may be a dose threshold in triggering the FLASH effect (37-40), Kang et al. (22) proposed the dose threshold dose rate (DTDR) that accounts for the individual spot dose rate and dose threshold. The dose is at its maximum at the spot center and decreases from the center to the lateral direction. DTDR uses a dose threshold to exclude the low dose tails of spots from dose rate calculation. Different dose rate quantifications may give rise to different treatment planning considerations for OAR dose rate assessment and machine delivery requirements. More biological studies are expected to determine the contributing factors in the PBS FLASH effect. Awareness of the differences in proton PBS dose rate calculation is important to design experiments and clinical trials to uncover FLASH-PT's biological and physiological mechanisms.

'FLASHness' can be evaluated by the dose rate coverage in OARs. The dose rate distribution can be calculated for each field then overlaid on the CT images. Similar to the dose-volume histogram (DVH) representation of a 3D dose distribution, the 3D dose rate distribution is concentrated using a single dose rate volume histogram (DRVH) curve to represent the voxel-based dose rate distribution (22). By specifying a FLASH dose rate threshold of 40 Gy/s, a dose rate coverage index $V_{40Gy/s}$ was defined, representing the percentage of the volume receiving a dose rate >40 Gy/s.

**PBS Transmission FLASH-PT**

Several research groups have explored PBS transmission FLASH treatment planning (17, 22-26) based on theoretical assumptions, and biological evidence. Overall, these groups have proposed their ways of quantifying dose rate in a treatment plan using metrics as described previously. In general, the following conditions are assumed: 1) the relative biological effect (RBE) for both FLASH and non-FLASH irradiated volumes are both 1.1; 2) Multiple fields do not change the FLASH effect if single beams meet the FLASH dose rate, which is usually > 40 Gy/s; 3) There is no dose threshold for the FLASH sparing effect.

The FLASH-PT plan quality includes dose uniformity in target, doses to organs-at-risks (OARs), and dose rates distributions. Besides, how transmission plans compare to conventional Bragg peak plans in plan quality also needs to be identified and characterized. van de Water et al. (17) first discussed evaluating spatially varying instantaneous dose rates for different intensity-modulated proton therapy (IMPT) planning strategies and delivery scenarios and comparing these with FLASH dose rates (>40 Gy/s) to four H&N cases using 2 Gy and 6 Gy fractions. The results suggested a number of factors should be required in order to achieve FLASH compatible dose rates, such as increased spot-wise beam intensities, spot-reduced planning, hypofractionation, etc. van Marlen et al. (25) evaluated the 244 MeV proton transmission plans by accounting for parameters relevant to dose rate and also compared the plan quality with the volumetric-modulated arc therapy (VMAT) plans, with a cohort of 7 lung cancer patients. They observed 1) the FLASH coverage of a scanning proton beam increases by increasing the max dose rate of a single spot. 2) 100% FLASH coverage cannot be achieved due to distal spots with a low dose rate. 3) Superior or equal plan quality in lungs, thoracic wall, and heart using transmission beams compared to VMAT plans. Similarly, they applied their FLASH dose rate quantification method to an H&N study of 10 patients (23) and compared the transmission plans to both conventional IMPT and VMAT plans. Results suggested comparable OAR sparing of transmission beams to conventional IMPT and better than VMAT.

As PBS FLASH machine delivery has become more attainable with high beam current achieved in major vendors, it becomes interesting to directly apply transmission planning based on the realistic parameters from the delivery systems and other dose rate quantification methods. Kang et al. (22) studied the transmission plans using 240 MeV proton beams by applying a realistic machine setting based on the Varian ProBeam system for lung cancer patients. By using two different minimum MUs/spot of 100 and 400, the average plan quality of the lung plans is also characterized, with the 100 minimum MU/spot case achieving slightly better results. Using a minimum MU/spot of 400 also achieves better FLASH coverage in OARs according to three PBS dose rate quantitation methods.

Another important factor in treatment planning is the number of beams and beam arrangement. Wei et al. (41) studied the FLASH treatment planning for hypofractionation liver cancer cases by considering machine-related factors including beam current, minimum MU/spot, spot reduction, minimum spot time, and realistic plan-related factors including beam angular arrangement and number of fields. The results suggest minimum spot time and the number of beams are important parameters for the liver plans' FLASH dose rate coverage, target uniformity and coverage, and OAR dose constraints. The FLASH dose rate coverage can be improved by reducing either the minimum spot time or the number of beams. Reducing the minimum spot time alone can also potentially improve the plan quality as the minimum MU constraint is less strict.

One recent proton FLASH treatment planning study (24) assumed a normal tissue protection factor for transmission FLASH beams in H&N cases, where relative biological effective (RBE) dose is halved when irradiated with FLASH dose rate. The resulted dose volume histograms from the transmission plans indicate a significant reduction of H&N dose metrics, corresponding to the FLASH sparing effect.

**PBS Bragg peak FLASH-PT**

High-energy transmission beams are associated with high beam current and high delivery robustness regarding range uncertainty, which offers excellent advantages in FLASH treatment planning and application. Nonetheless, there are still several disadvantages. 1) The transmission beams do not use the high LET region of the Bragg peak, which may limit the cell-killing capability in the target. 2) They cannot spare the normal tissue beyond the target in the beam paths. 3) They cannot shoot through at certain angles and may affect the overall plan quality, especially for relatively large patients, where, despite using the highest cyclotron energy 250 MeV, the Bragg peak may still stop in the patient body. Accounting for the deficiency of the transmission beams, Lin et al. proposed a combined dose and dose rate optimization method that places the non-FLASH Bragg peak beams in the tumor to improve target dose conformity and spare normal tissue and uses the FLASH transmission beams to cover the tumor boundary to achieve high dose rate coverage at the OARs (42).

To fully eliminate the exit dose of transmission beams, a single-energy Bragg peak FLASH delivery method was developed and can offer excellent dosimetry characteristics (12). The first proof-of-concept study has demonstrated the feasibility of using a single-energy Bragg peak for FLASH treatment (28). The successful implementation of Bragg peak FLASH relies on the inverse TPS, hardware (universal range shifters (URS) and range compensators (RC)), and available proton beam current in the treatment room. The URS and RC systems can pull back the range of the high-energy proton beams to align the Bragg peak with the target exit edge, which shows promising results in achieving comparable ultra-high OAR dose rate and superior dose quality compared to the proton transmission plans. The method was then thoroughly evaluated in hypofractionated lung and liver dosimetry and dose rate study by accounting for practical clinical beam arrangement and fractionations (41,43) (Figures 3 and 4). These studies indicate that the Bragg peak FLASH-PT planning can reach a sufficiently high dose rate with feasible beam currents and beam arrangement despite different dose rate calculation methods for both lung and liver cancers. The advancement of clinical transitions of proton Bragg peak FLASH-PT may offer optimal treatment solutions for cancer patients.

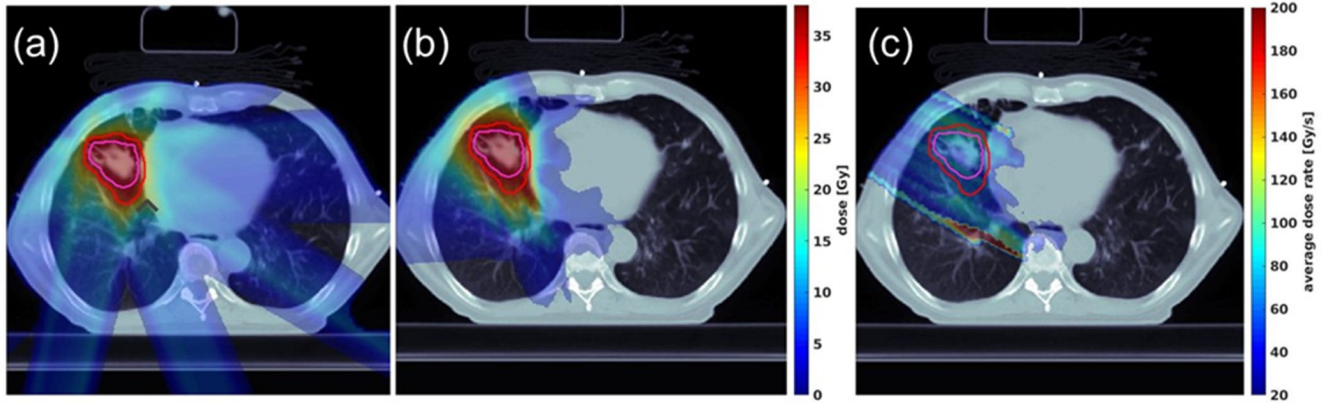

Fig. 3 (a) A typical FLASH proton PBS transmission plan (b) A typical single-energy FLASH proton PBS Bragg peak plan for a lung patient, with a dose prescription of 18 Gy/fraction in the target. (c) The average dose rate distribution of one single-energy proton Bragg peak beam indicating > 40 Gy/s ADR is reached in most normal tissue.

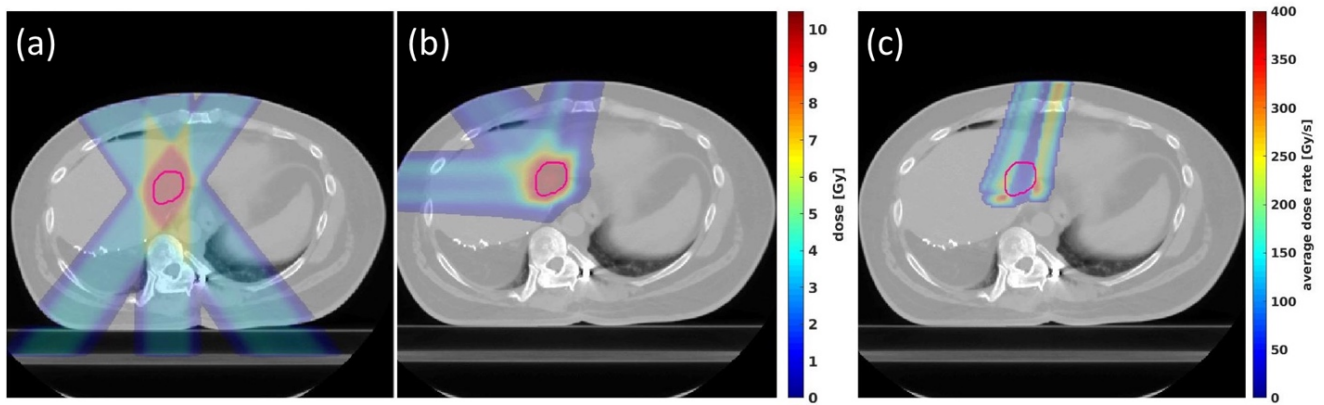

Fig. 4 (a) A typical FLASH proton PBS transmission plan (b) A typical single-energy proton PBS Bragg peak FLASH plan for a liver patient, with a dose prescription of 10 Gy/fraction. (c) The average dose rate distribution of one single-energy PBS proton Bragg peak beam indicating > 40 Gy/s ADR is reached in most normal tissue.

**Biological Investigations**

Most of the current biological studies studying the FLASH effect have used photon, electron, and scattered proton beams. In the first biological study using PBS proton beams, Cunningham et al. (10) delivered a uniform physical dose of 35 Gy (toxicity study) or 15 Gy (tumor control study) to the right hind leg of mice at various dose rates of 1, 57 and 115 Gy/s, using the plateau region of a 250 MeV proton beam, and quantified acute and delayed radiation effects using varied surrogates including skin and plasma levels of different cytokines, skin toxicity, and hind leg contractures. The results indicated comparable tumor control between conventional and FLASH dose rates. Plasma and skin levels of TGF-β1, skin toxicity, and leg contracture were significantly decreased in FLASH compared to conventional groups of mice. In this study, the average dose rate defined by Folkerts et al. (36) was used in the PBS fields, which is a conservative yet effective way in calculating the dose rates as 57 Gy/s established as triggering the FLASH effect.

The first human clinical trial using FLASH PBS transmission beams ("FAST-01") (27) was initialized, and up to 10 patients with bone metastases are expected to enroll to evaluate clinical workflow feasibility, treatment-related side effects, and efficacy of treatment as assessed by measuring pain relief of trial participants.

Currently, there is no established explanation for the FLASH effect, although several hypotheses have been proposed, and evidence has been reported to support mechanisms including rapid oxygen depletion (37, 44-47), reactive oxygen species (48-49), and immune response (50-52). There have been a number of follow-up studies based on the first principle of oxygen depletion models that have been applied to electron (53), photon (54), and proton beams (55). For PBS proton beams, due to the scanning pattern of a field, the dose-time profile for each voxel in a field may be unique; therefore, it may be interesting to investigate the oxygen depletion that combines the PBS. Rothwell et al. (55) proposed a tool that explores oxygen depletion. In particular, they combined the spatial and temporal information in the PBS delivery, including spot scanning, energy layer switching, and gantry rotation, to their proposed oxygen depletion model that accounts for simplified oxygen diffusion in blood vessels. They also proposed novel metrics such as dose- oxygen enhancement ratio (OER) histograms to evaluate the radiobiological effect of selected regions in a field.

**Discussion and Conclusion**

The recent progress of proton PBS FLASH-PT, including the novel PBS Bragg peak FLASH technique, is promising. The existing studies have identified the key factors, providing guidance in machine design for transmission and single-energy Bragg peak beams to serve the clinical application need, corroborated with pioneering biological data supporting the PBS FLASH-PT using mice models. Major challenges for conventional proton plan delivery, including motion management and robustness (56-58), have not been well covered based on existing literature. PBS FLASH beams are expected to be delivered very quickly in much less than a second, which effectively "freezes" the tissue motion. This poses a separate challenge also opportunity in delivering the PBS FLASH beam to the target in a precise phase of organ motion. As Bragg peak beams are inherently more sensitive to range uncertainties compared to transmission beams, separate studies detailing robust characteristics and the role of robustness optimization planning for Bragg peak beams are needed. In addition, potential interruptions of beam delivery could pose challenges to the clinical implementation of PBS FLASH and disrupt the FLASH effect. The potentials of Bragg peak planning warrant separate studies in other treatment sites regarding the differences in fractionated dose prescriptions, patient anatomy, specific constraints, and dose rate distribution, etc.

Due to machine delivery limits and challenges such as energy layer switch and available beam currents in the treatment room, both the single-energy transmission and Bragg peak beams are practicable for proton PBS FLASH applications. Also, with the FLASH biological effect remaining undetermined, it is prudent to assume conservative constraints for proton FLASH PBS treatment plans. The future development in the proton PBS machines will improve beam transmission efficiency from accelerator to treatment room, enabling more freedom in PBS FLASH-RT and facilitating faster translation from research to clinics.

**Acknowledgments**

The authors would like to acknowledge the useful discussions with Dr. Michael Folkerts from Varian.

**Funding:** This research was funded in part through the NIH/NCI Cancer Center Support Grant (No. P30 CA008748).

**References**

1. Favaudon V, Caplier L, Monceau V, et al. Ultrahigh dose-rate FLASH irradiation increases the differential response between normal and tumor ti1010ssue in mice. Sci Transl Med 2014;6. https://doi. org/10.1126/scitranslmed.3008973. 245ra93-245ra93
2. Montay-Gruel P, Acharya MM, Petersson K, et al. Long-term neurocognitive benefits of FLASH radiotherapy driven by reduced reactive oxygen species. PNAS 2019;116:10943–51. https://doi.org/ 10.1073/pnas.1901777116
3. Vozenin M-C, De Fornel P, Petersson K, et al. The advantage of FLASH radiotherapy confirmed in mini-pig and cat-cancer patients. Clin Cancer Res 2018. https://doi.org/10.1158/1078-0432.CCR-17-3375
4. Zlobinskaya O, Siebenwirth C, Greubel C, et al. The effects of ultra-high dose rate proton irradiation on growth delay in the treatment of human tumor xenografts in nude mice. Radiat Res 2014;181:177–83. https://doi.org/10.1667/RR13464.1
5. Diffenderfer ES, Verginadis II, Kim MM, et al. Design, implementation, and in vivo validation of a novel proton FLASH radiation therapy system. Int J Radiat Oncol Biol Phys 2020;106:440–8. https:// doi.org/10.1016/j.ijrobp.2019.10.049
6. Loo BW, Schuler E, Lartey FM, et al. (P003) delivery of ultra-rapid flash radiation therapy and demonstration of normal tissue sparing after abdominal irradiation of mice. Int J Radiat Oncol Biol Phys 2017;98:E16. https://doi.org/10.1016/j.ijrobp.2017.02.101
7. Simmons DA, Lartey FM, Schüler E, et al. Reduced cognitive deficits after FLASH irradiation of whole mouse brain are associated with less hippocampal dendritic spine loss and neuroinflammation. Radiother Oncol 2019;139:4–10. doi: 10.1016/j.radonc.2019.06.006
8. Levy K, Natarajan S, Wang J, et al. FLASH irradiation enhances the therapeutic index of abdominal radiotherapy in mice. bioRxiv 2019
9. Levy K, Natarajan S, Wang J. et al. Abdominal FLASH irradiation reduces radiation-induced gastrointestinal toxicity for the treatment of ovarian cancer in mice. Sci Rep 2020. 10: 21600
10. Cunningham S, McCauley S, Vairamani K, et al. FLASH proton pencil beam scanning irradiation minimizes radiation-induced leg contracture and skin toxicity in mice. Cancers 2021, 13, 1012
11. Bourhis J, Sozzi WJ, Jorge PG, et al. Treatment of a first patient with FLASH-radiotherapy. Radiother Oncol 2019;139:18–22. https://doi.org/10.1016/j.radonc.2019.06.019
12. Chow R, Kang M, Wei S, et al. FLASH Radiation Therapy: Review of the Literature and Considerations for Future Research and Proton Therapy FLASH Trials. Appl Radiat Oncol. 2021;10(2):16-21

13. Esplen N, Mendonca MS, Bazalova-Carter M. Physics and biology of ultrahigh dose-rate (FLASH) radiotherapy: A topical review. *Phys. Med. Biol.* 2020;65:23TR03
14. Zou W, Diffenderfer ES, Cengel KA, et al. Current Delivery Limitations of Proton PBS for FLASH. Radiotherapy and Oncology 2021; 155 :212–21
15. Koschik A, Bula C, Duppich, J, et al. GANTRY 3: Future development of the PSI PROSCAN proton therapy facility. In Proceedings of the 6th International Particle Accelerator Conference, IPAC2015, Richmond, VA, USA, 3–8 May 2015
16. Shen J, Tryggestad E, Younkin J, et al. Using experimentally determined proton spot scanning timing parameters to accurately model beam delivery time. *Med. Phys*. **2017**, *44*, 5081–5088
17. van de Water S, Safai S, Schippers JM, et al. Towards FLASH proton therapy: the impact of treatment planning and machine characteristics on achievable dose rates. Acta Oncol 2019;58:1463–9. https://doi.org/ 10.1080/0284186X.2019.1627416
18. Kourkafas G, Bundesmann J, Fanselow T, et al. FLASH proton irradiation setup with a modulator wheel for a single mouse eye. Med. Phys 2021, 48, 1839–1845
19. Kim, MM, Verginadis II, Goia D, et al. Comparison of FLASH Proton Entrance and the Spread-Out Bragg Peak Dose Regions in the Sparing of Mouse Intestinal Crypts and in a Pancreatic Tumor Model. *Cancers* 2021(13): 4244
20. Patriarca A, Fouillade C, Auger M, et al. Experimental set-up for FLASH proton irradiation of small animals using a clinical system. Int J Radiat Oncol Biol Phys 2018;102:619–26. https://doi.org/ 10.1016/j.ijrobp.2018.06.403
21. Evans T, Cooley J, Wanger M, et al. Demonstration of the FLASH effect within the Spread-out Bragg Peak after abdominal irradiation of mice. IJPT 2021, doi 10.14338/IJPT-20-00095
22. Kang M, Wei S, Choi JI, et al. Quantitative assessment of 3D dose rate for proton PBS FLASH RT and its application in a deliverable system for Lung hypofractionation treatment planning, Cancers 2021, 13(14): 3549
23. van Marlen P, Dahele M, Folkerts M, et al. Ultra-High Dose Rate Transmission Beam Proton Therapy for Conventionally Fractionated Head and Neck Cancer: Treatment Planning and Dose Rate Distributions. Cancers 2021. 13(8):1859
24. Verhaegen F, Wanders RG, Wolfs C, et al. Considerations for shoot-through FLASH proton therapy. Phys Med Biol 2021. 66(6):06NT01. doi: 10.1088/1361-6560/abe55a
25. van Marlen P, Dahele M, Folkerts M, et al. Bringing FLASH to the clinic: treatment planning considerations for ultrahigh dose-rate proton beams. Int J Radiat Oncol Biol Phys 2020;106:621–9
26. Gao H, Lin B, Lin Y, et al. Simultaneous dose and dose rate optimization (SDDRO) for FLASH proton therapy. Med Phys 2020

27. First Patient Treated in FAST-01 FLASH Therapy Trial. Available online: **https://www.appliedradiology.com/articles/varian-first-patient-treated-in-fast-01-flash-therapy-trial**
28. Kang M, Wei S, Choi JI, et al. A universal range shifter and range compensator can enable proton pencil beam scanning single-energy Bragg peak FLASH-RT treatment using current commercially available proton systems, Int J Radiat Oncol Biol Phys. 2022 May 1;113(1):203-213
29. Jolly, S.; Owen, H.; Schippers, M.; Welsch, C. Technical challenges for FLASH proton therapy. Phys. Med. 2020, 78, 71–82

30. Nesteruk KP, Togno M, Grossmann M, et al. Commissioning of a clinical pencil beam scanning proton therapy unit for ultrahigh dose rates (FLASH). Med Phys 2021;48(7):4017-4026
31. Chang C, Huang S, Harms J, et al. A standardized commissioning framework of Monte Carlo dose calculation algorithms for proton pencil beam scanning treatment planning systems. Med Phys 2020; https://doi.org/10.1002/mp.14021
32. Nesteruk, K. P.; Psoroulas, S. FLASH Irradiation with Proton Beams: Beam characteristics and their implications for beam diagnostics. Appl. Sci. 2021, 11(5), 2170
33. Srinivasan S, Duperrex PA, Schippers JM. Beamline characterization of a dielectric-filled reentrant cavity resonator as beam current monitor for a medical cyclotron facility. Phys Med. 2020 Oct;78:101-108. doi: 10.1016/j.ejmp.2020.09.006. Epub 2020 Sep 18. PMID: 32956916
34. Dölling, R. Ionisation Chambers and Secondary Emission Monitors at the PROSCAN Beam Lines. AIP Conf. Proc. 2006, 868, 271–280
35. Srinivasan S, Duperrex PA, Schippers JM. Beamline characterization of a dielectric-filled reentrant cavity resonator as beam current monitor for a medical cyclotron facility. Phys. Med. 2020, 78, 101–108
36. Folkerts M, Abel E, Busold S et al. A Framework for defining FLASH dose rate for pencil beam scanning, Medical physics 2020; https://doi.org/10.1002/mp.14456
37. Adrian G, Konradsson E, Lempart M, et al. The FLASH effect depends on oxygen concentration. *Br J Radiol* 2020;93:20190702
38. Wilson P, Jones B, Yokoi T, et al. Revisiting the ultra-high dose rate effect: Implications for charged particle radiotherapy using protons and light ions. *Br J Radiol.* 2012;85:e933–e939
39. Bourhis J, Montay-Gruel P, Jorge PG, et al. Clinical translation of FLASH radiotherapy: Why and how? *Radiother Oncol* 2019;139:11–17
40. Krieger M, van de Water S, Folkerts MM, et al. A quantitative FLASH effectiveness model to reveal potentials and pitfalls of high dose rate proton therapy. Med Phys. 2022 Mar;49(3):2026-2038
41. Wei S, Lin H, Choi JI, et al. FLASH radiotherapy using single-energy proton PBS transmission beams for hypofractionation liver cancer: dose and dose rate quantification, Frontiers in Oncology, Front Oncol. 2022 Jan 13;11:813063
42. Lin Y, Lin B, Fu J, et al. SDDRO-Joint: simultaneous dose and dose rate optimization with the joint use of transmission beams and Bragg peaks for FLASH proton therapy Phys Med Biol 66 125011
43. Wei S, Lin H, Choi JI, et al. Use single-energy proton pencil beam scanning Bragg peak for intensity-modulated proton therapy FLASH treatment planning in liver hypofractionation radiotherapy. Under review, 2022
44. Pratx G, Kapp DS. A computational model of radiolytic oxygen depletion during FLASH irradiation and its effect on the oxygen enhancement ratio. Phys Med Biol 2019. 64 185005
45. Rothwell BC, Kirkby NF, Merchant MJ, et al. Determining the parameter space for effective oxygen depletion for FLASH radiation therapy. Phys Med Biol. 2021; 66: 055020
46. Petersson K, Adrian G, Butterworth K, et al. A Quantitative Analysis of the Role of Oxygen Tension in FLASH Radiation Therapy. Int J Radiat Oncol Biol Phys 2020, 107, 539–547
47. Zhou S, Zheng D, Fan Q, et al. 2020 Minimum dose rate estimation for pulsed FLASH radiotherapy: a dimensional analysis, Med Phys 47 3243–9

48. Labarbe R, Hotoiu L, Barbier J, et al. A physicochemical model of reaction kinetics supports peroxyl radical recombination as the main determinant of the FLASH effect. *Radiother Oncol* 2020;153:303-310
49. Spitz DR, Buettner GR, Petronek MS, et al. An integrated physico-chemical approach for explaining the differential impact of FLASH versus conventional dose rate irradiation on cancer and normal tissue responses. Radiother Oncol 2019, 139:23-27
50. Buonanno M, Grilj V, Brenner DJ. Biological effects in normal cells exposed to FLASH dose rate protons. Radiother Oncol 2019 Oct; 139():51-55
51. Durante M, Bräuer-Krisch E, Hill M. Faster and safer? FLASH ultra-high dose rate in radiotherapy. Br J Radiol 2018 Feb; 91(1082):20170628
52. Jin JY, Gu A, Wang W, et al. Ultra-high dose rate effect on circulating immune cells: A potential mechanism for FLASH effect? Radiother Oncol 2020;149:55-62
53. Khan S, Bassenne M, Wang J, et al. Multicellular Spheroids as In Vitro Models of Oxygen Depletion During FLASH Irradiation. Int J Radiat Oncol Biol Phys 2021;110(3):833-844
54. Lyu Q, Neph R, O'Connor D, et al. ROAD: ROtational direct Aperture optimization with a Decoupled ring-collimator for FLASH radiotherapy. Phys Med Biol 2021. 29;66(3):035020
55. Rothwell BC, Lowe M, Kirkby NF, et al. Oxygen Depletion in Proton Spot Scanning: A Tool for Exploring the Conditions Needed for FLASH. *Radiation* 2021; 1(4):290-304
56. Lin L, Kang M, Huang S, et al. Beam-specific planning target volumes incorporating 4D CT for pencil beam scanning proton therapy of thoracic tumors. J Appl Clin Med Phys. 2015;16(6):5678
57. Kang M, Huang S, Solberg TD, et al. A study of the beam-specific interplay effect in proton pencil beam scanning delivery in lung cancer. Acta Oncol 2017;56(4):531-540
58. Lin L, Souris K, Kang M, et al. Evaluation of motion mitigation using abdominal compression in the clinical implementation of pencil beam scanning proton therapy of liver tumors. Med Phys 2017;44 (2)